\documentclass[a4paper]{llncs}
\usepackage[parfill]{parskip}    
\usepackage[english]{babel}
\usepackage{amsmath}
\usepackage{amssymb}
\usepackage{graphicx}
\usepackage[dvipsnames]{xcolor}
\usepackage{tikzscale}
\usepackage{tikz}
\usepackage[normalem]{ulem}
\usetikzlibrary{positioning, arrows, shapes.misc, calc}

\usepackage{tabularx}
\usepackage{algorithm}
\usepackage{algpseudocode}

\usepackage[colorinlistoftodos]{todonotes}

\usepackage{enumitem}
\newlist{indenteddesc}{description}{1}
\setlist[indenteddesc]{
  leftmargin=5em,  
  rightmargin=3em,
  labelindent=3em, 
  labelwidth=1.5em,
  labelsep=.5em
}

\usepackage{acronym}

\title{
Statistical abstraction for multi-scale spatio-temporal systems
\thanks{M.M. and G.S.\ are supported by the European Research Council under grant MLCS 306999. J.H.\ is supported by the EU project, QUANTICOL 600708.}}
\author{
Michalis Michaelides\inst1
\and\\ Jane Hillston\inst1
\and Guido Sanguinetti\inst1\inst2
}
\institute{School of Informatics, University of Edinburgh \and SynthSys, Centre for Synthetic and Systems Biology, University of Edinburgh}

\begin{document}

\acrodef{CTMC}{Continuous Time Markov Chain}
\acrodef{GP}{Gaussian Process}
\acrodef{JSD}{Jensen-Shannon Divergence}

\maketitle

\begin{abstract} \label{abstract}
Spatio-temporal systems exhibiting multi-scale behaviour are common in applications ranging from cyber-physical systems to systems biology, yet they present formidable challenges for computational modelling and analysis. Here we consider a prototypic scenario where spatially distributed agents decide their movement based on external inputs and a fast-equilibrating internal computation. We propose a generally applicable strategy based on statistically abstracting the internal system using Gaussian Processes, a powerful class of non-parametric regression techniques from Bayesian Machine Learning. We show on a running example of bacterial chemotaxis that this approach leads to accurate and much faster simulations in a variety of scenarios. 
\end{abstract}

\section{Introduction}
Modelling spatially extended dynamical systems is a task of central importance in science and engineering. Examples range from cyber-physical systems, to collective adaptive systems of human behaviour, to cellular systems. Despite their importance, computational modelling and analysis of such systems remains challenging due to a number of factors: the large number of degrees of freedom, the intrinsically hybrid nature of discrete systems existing in  continuous space, and, frequently, the existence of multiple temporal scales in the system. As a result of these features, computational simulation of such systems is generally onerous, particularly in a stochastic setting \cite{gilbert_multiscale_2015,dada_multi-scale_2011}.

In this paper, we consider the scenario where the system consists of multiple, spatially distributed, identical agents. The agents can sense an external, deterministic field and use this information to perform a stochastic, internal computation which determines the agent's subsequent move. The internal computation is often a system which will quickly reach a steady-state equilibrium when left unperturbed, e.g. a chemical reaction network. While this scenario is a special case as the agents do not interact with each other, it is sufficiently generic to cover many application scenarios, such as autonomous drones performing a task in space, or bacteria exploring a nutrient field. Such systems are cumbersome to handle computationally as the simulation of the internal computation needs to be repeated at every spatial step, so that simulating a single trajectory of the overall system may involve hundreds of simulations of the internal model.

Here we propose a novel approach to alleviate this computational burden based on emulating the statistics of the internal system. The central idea is to replace the expensive computation of the internal system with a lookup table which maps external stimulus to the output behaviour of the internal system. Crucially, we do not aim to model the detail of the internal state, but only an abstracted version capturing its qualitative behaviour (formalised as a logical property satisfied by the states). We achieve this by learning a parameters-to-behaviours regression map using Gaussian Processes (GPs), a powerful class of non-parametric Bayesian regression models. Our work is motivated by earlier work on using GPs to learn effective characterisations of system behaviour \cite{bortolussi_smoothed_2016,bortolussi_efficient_2015,michaelides_property-driven_2016}.

The rest of the paper is organised as follows: background on spatio-temporal systems and \emph{E.\ coli} chemotaxis which serves as a running example (Section \ref{sec:bg}); the general framework for our statistical abstraction methodology, and its application to the chemotaxis system (Section \ref{sec:method}); results assessing the quality and efficiency of the abstraction (Section \ref{sec:results}); closing remarks about prospective expansion of the work (Section \ref{sec:conclusion}).

\section{Background}\label{sec:bg}

\subsection{Spatio-temporal agent models}
We start by defining the class of spatio-temporal agent models we will consider  in this paper. Let $\mathcal{D}$ be a spatial domain (usually a compact subset of $\mathbb{R}^n$ with $n=2,3$), and let $[0,T]$ be the temporal interval of interest. We define the {\it spatio-temporal field}  $f\colon\mathcal{D}\times[0,T]\to\mathbb{R}$ to be a real-valued function defined on the spatial and temporal domains of interest.
A spatio-temporal agent model is a triple $(\mathcal{D},f,\mathcal{A})$ where $\mathcal{A}$ is a collection of point {\it agents} whose location follows a stochastic process which depends on the spatio-temporal field. We note that this is  not the most general case, as agents may be spatially extended, interact with each other or even  influence the evolution of the spatio-temporal field. Nevertheless, such a level of abstraction is frequently adopted and justifiable  in many practical applications.

\paragraph{Running example: chemotaxis in the \emph{Escherichia coli} bacterium} %
\label{sec:bg:chemotaxis}
Foraging is a central problem for microbial populations. The bacterium \emph{Escherichia coli} will normally perform a random walk within a spatial domain where nutrient concentration is constant (e.g.\ a Petri dish). When presented with a spatially varying nutrient field, a phenomenon known as {\it chemotaxis} arises. As the bacterium performs a random walk in the nutrient field encountering changing nutrient levels, its sensory pathway effectively evaluates a temporal gradient of the nutrients (or ligands) it experiences; the walk is biased so that the bacterium experiences a positive temporal gradient more often than not \cite{vladimirov_dependence_2008,sourjik_responding_2012}. Since the bacterium is moving in the field, the temporal gradient is implicitly translated into a spatial one, so the bacterium drifts toward advantageous concentrations. 
Implicitly translating a temporal gradient to a spatial one through motion is necessary for the bacterium cell, because its body size is too small to allow for effective calculation of the spatial gradient of a chemical field at its location. As a result, we can safely regard the bacteria as point-like agents.

\subsection{Multi-scale models}
In many practical situations, one is interested in modelling not only the movement of the agents, but also the mechanism through which sensing and decision making is carried out within each agent. This naturally leads to structured models with distinct layers of organisation, with behaviour in each layer informing the simulation that takes place at the layer above or below.  We will assume that the internal workings of the agent are also stochastic, and we will model them as a {\it population Continuous Time Markov Chain} (pCTMC) \footnote {The pCTMC is the internal model for a \emph{single} agent here, not for multiple agents.}. Formally, a pCTMC is defined as follows.
\begin{definition}

A population CTMC is a continuous-time Markov chain \cite{norris_markov_1998} with a discrete state-space $\mathcal{X}$, and an associated transition rate matrix $Q$. Each state in $\mathcal{X}$ counts the number of entities of each type or ``species" in a population, $\mathbf{X}\in \left\{\mathbb{N}^0\right\}^d$ for $d$ species. Transitions in this space occur according to the rates given by $Q$.

The transitions can be regarded as occurrences of chemical reactions, written as
\begin{align}
\sum_{i=1}^d r_i X_i \xrightarrow{\tau(\mathbf{X})} \sum_{i=1}^d s_i X_i,
\end{align}
where for every species $X_i$, $r_i$ particles of $X_i$ are consumed and $s_i$ particles are created. The transition rate $\tau(\mathbf{X})$ depends upon the current state of the system, and is the rate parameter of an exponential distribution governing the waiting times for this transition. The above transition rates of allowed reactions reconstruct the rate matrix $Q$.

\end{definition}

\begin{figure}[!t]
	\centering
	\includegraphics[height=0.19\textwidth]{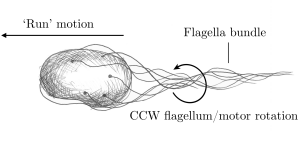}
	\includegraphics[height=0.19\textwidth]{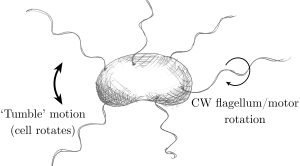}
	\includegraphics[height=0.2\textwidth]{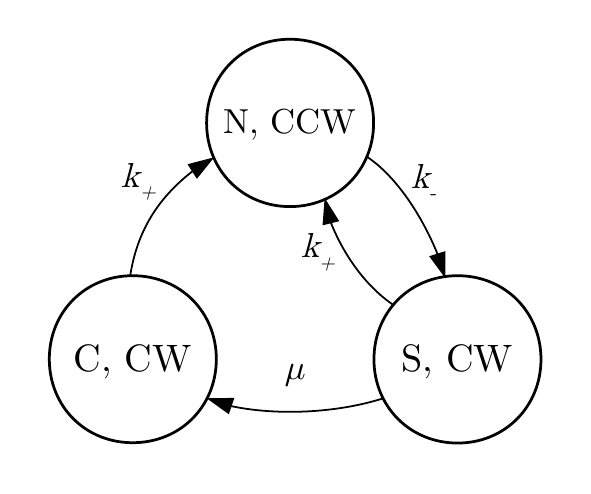}
	\caption{The two motility modes of an \emph{E.\ coli} cell. Left: the F/M are in CCW conformations, forming a helical bundle and propelling the cell. Centre: the F/M are in CW conformations, breaking the bundle apart and causing the cell to re-orient in place. Right: CTMC for a single F/M, with three conformation states and transition rates $k_\pm(m, L)$ and fixed $\mu = 5\text{s}^{-1}$.}
	\label{fig:CTMC_FM}
\end{figure}

\paragraph{Motor control in \emph{E.\ coli}}
An \emph{E.\ coli} cell achieves motility by operating \emph{multiple} flagellum/motor pairs (F/M), which can either drive it straight (subject to small Brownian perturbation), or rotate it in place. Thus, the cell can either be `tumbling' (re-orienting itself while stationary) or `running' (propelling itself forward while maintaining direction) at any time (Figure \ref{fig:CTMC_FM}: left, centre). The motility state, RUN/TUMBLE, of the cell is determined by the number of flagella found in particular conformations. The model in \cite{sneddon_stochastic_2012} suggests three possible conformations for a flagellum: \emph{curly} ($C$), \emph{semicoiled} ($S$) and \emph{normal} ($N$). The associated motor is modelled as a stochastic bistable system, which rotates either clockwise (CW) or counter-clockwise (CCW). Changes in motor rotation induce conformational changes on the associated flagellum. Transition rates between motor states are given by rate parameters $k_+$ and $k_-$ for transitions CW $\to$ CCW and CCW $\to$ CW, respectively.  The possible transitions between  flagellum/ motor states are summarised in the schematic diagram in Figure \ref{fig:CTMC_FM}: right.  {\it E.\ coli} normally has of the order of ten flagella and associated motors; the dynamics of the pair flagellum/ motor population therefore lends itself to be easily described as a pCTMC.
The $k_\pm$ transition rates depend on the temporal gradient evaluated by the chemotaxis pathway, and represent the functional interface of the bacterium with its external environment.


The classical mathematical model for the sensory response of the cell to external ligand concentration changes is provided by the Monod-Wyman-Changeux (MWC) model   \cite{sourjik_functional_2004,sneddon_stochastic_2012,hansen_chemotaxis_2008}. The model considers sensor clusters which signal information about ligand concentration changes to the motors, by triggering a biochemical response in the cell (phosphorylation of the CheY protein which binds to the motors) affecting the switching rates of rotation direction, $k_\pm$.


The full MWC model is still highly complex; in practice, we follow \cite{sneddon_stochastic_2012} and adopt a simplified model of sensory response to describe the dependency of motor rates $k_{\pm}$ on ligand concentrations. This consists in abstracting the CheY signalling pathway in an effective variable $m$, which represents the methylation state of the ligand receptors and whose stochastic evolution is dependent on the ligand concentration $L$. Since $m$ depends on past $L$ concentrations the cell has been in, one may think of it as a \emph{chemical memory} of sorts which encodes the value of $L$ at previous times. The time comparison window is determined by how fast methylation happens --- faster methylation leads to a shorter memory.

Sneddon et al \cite{sneddon_stochastic_2012} then resolve the entire dependency chain of the chemotaxis pathway to Equations \ref{eq:motor_rate} and \ref{eq:meth}.
The motor switching rates $k_\pm(m, L)$ are given by the deterministic equation
\begin{align}
\label{eq:motor_rate}
k_{\pm} = &\omega \cdot \exp\bigg\{ \pm \bigg[ \frac{g_0}{4} - \frac{g_1}{2}
									\bigg( \frac{Y_p(m, L)}{Y_p(m, L) + K_D}
									\bigg)
							\bigg] \bigg\},
\end{align}
where
\begin{align*}
Y_p(m, L) = &\alpha \cdot \left[ 1 + e^{\epsilon_0 + \epsilon_1 m} \cdot 
	\left(\frac{1 + L / K_\text{TAR}^\text{off}}
				{1 + L / K_\text{TAR}^\text{on}}\right) ^ {n_\text{TAR}}
	\cdot
	\left(\frac{1 + L / K_\text{TSR}^\text{off}}
				{1 + L / K_\text{TAR}^\text{on}}\right) ^ {n_\text{TSR}}
\right]^{-1}.
\end{align*}
The methylation process can be naturally modelled as a birth~/~death process with rates depending on ligand concentration; again following \cite{sneddon_stochastic_2012}
we take a fluid approximation of this, yielding the Ornstein-Uhlenbeck (OU) process:
\begin{align}
\label{eq:meth}
\frac{dm}{dt} = -\frac{1}{\tau}(m - m_0(L)) + \eta_m(t).
\end{align}
In the above stochastic differential equation (SDE), $\eta_m = \sigma_m \sqrt{2/\tau} \Gamma(t)$, $\Gamma(t)$ is the normally distributed random process with 0 mean and unit variance, $\sigma_m$ is the standard deviation of fluctuations in the methylation level, and $m_0(L)$ is an empirically derived function whose output is the methylation level required for full adaptation at the current external ligand concentration $L$.  The adaptation rate $\tau$, determines how fast methylation occurs and so, how long the `chemical memory' of previous $L$ values is in the system. The constants $\tau$, along with $mb_0$ and $\alpha$ involved in the $m_0(L)$ function (see \cite{sneddon_stochastic_2012}), fully parametrise the methylation evolution.
See \cite{vladimirov_predicted_2010} for reported values of constants used in Equation \ref{eq:motor_rate} and \cite{frankel_adaptability_2014,sneddon_stochastic_2012} for a detailed derivation of the results.
Equations \ref{eq:motor_rate} and \ref{eq:meth} couple the transition rates of the pCTMC in Figure~\ref{fig:CTMC_FM}:~Right, with the external ligand concentrations, and therefore fully describe the internal model of the {\it E.\ coli} chemotactic response.


\subsection{Simulating multi-scale systems}
Multi-scale spatio-temporal systems are in general amenable to analytical techniques only in the simplest of cases. For the vast majority of real-world models, simulation-based analysis is the only option to gain behavioural insights.

Simulation of spatio-temporal systems typically employs nested algorithms: having chosen a time-discretisation for the spatial motion (which is assumed to have the slower time-scale), a spatial step is taken. Then, the value of the external field is updated, and the internal model is run for the duration of a given time-step with the new rates (corresponding to the updated value of the external field). A sample from the resulting state distribution then determines the velocity of the agent for the next time-step.

Clearly, this iterative procedure, while asymptotically exact (in the limit of small time discretisation), is computationally very demanding. This has motivated several lines of research in recent years \cite{bortolussi_efficient_2015,goutsias_quasiequilibrium_2005,rao_stochastic_2003,haseltine_approximate_2002}.

\paragraph{Simulating chemotaxis in {\it E.\ coli}}
Simulations of the {\it E.\ coli} model outlined previously proceed along the general lines discussed above. Given a value of the ligand field and a characteristic time-step $\Delta t$, we draw samples of the SDE \eqref{eq:meth} using the Euler-Maruyama method, a standard method for simulating SDEs.

{
In the F/M pCTMC system in the reaction equation style, each species represents a different F/M conformation for a total of three species. The following transitions occur:
\begin{align}
\label{eq:pCTMC1}
\begin{split}
(S\_CW) \xrightarrow{\mu} (C\_CW),~&~
(C\_CW) \xrightarrow{k_+} (N\_CCW), \\
(S\_CW) \xrightarrow{k_+} (N\_CCW),~&~
(N\_CCW) \xrightarrow{k_-} (S\_CW).
\end{split}
\end{align}
Note that in the above rate transitions there are dependencies on both external ($L$) and internal ($m$) states: $k_\pm(m, L)$, where $L$ is an external input to the system (the external chemoattractant concentration at the time) and $m$ is the current methylation level (sampled from the OU process in Equation \ref{eq:meth} every $\Delta t$). Instead, the rate transition for $(S\_CW)\to(C\_CW)$ is fixed, $\mu = 5\text{s}^{-1}$.
}



Using the exact Gillespie algorithm \cite{gillespie_exact_1977}, we then simulate the internal pCTMC for a length of time $\Delta t$ to draw a sample configuration of the flagella/ motor system. Formally, trajectories of length $\Delta t$ are checked against a property specifying the motility state for the cell (RUN/TUMBLE),
\begin{align}
\label{eq:phi_run}
\phi_\text{RUN}(\mathbf{s}) = (N \geq 2) \land (S = 0),
\end{align}
where $\mathbf{s} = (S, C, N)$ is the last state of the flagella/ motor pairs in the CTMC trajectory.

{
The spatial location of the bacterium is then updated according to a simple rule: if the sampled internal state corresponds to RUN, the agent moves rectilinearly and updates its position $\vec{r} \gets \vec{r} + \vec{v} \cdot \Delta t$, where $v=20\mu\text{m/s}$, the speed of the bacterium. Otherwise, if the internal state corresponds to TUMBLE, the agent remains still and its velocity is updated $\vec{v} \gets R(\theta) \cdot \vec{v}$, where $R(\theta)$ is the standard 2D unitary rotation matrix through an angle $\theta$, and $\theta$ is a tumbling angle sampled from a Gamma distribution as reported in Sneddon et al. \cite{sneddon_stochastic_2012}.
}

The above simulation scheme, outlined in Algorithm \ref{alg:sim-fine}, produces a chemotactic response to a ligand gradient. { It takes $\sim 270\text{s}$ to simulate a single cell trajectory of $t_\text{end} = 500\text{s}$ with a time-step $\Delta t=0.05$.}

\algnewcommand{\LineComment}[1]{\Comment{\parbox[t]{.45\linewidth}{#1}}}
\begin{algorithm}[!htb]
\caption{Simulation scheme for the \emph{E.\ coli} model, based on full simulation of the pCTMC describing F/M conformation changes. Below, $\tau$, $mb_0$, $\alpha$ are constants which parametrise the model (see \cite{sneddon_stochastic_2012}), and $\Delta t$ is the fixed simulation time-step.}
\label{alg:sim-fine}

\begin{algorithmic}[1]
\Function{Run}{$\vec{r}$, $\vec{v}$, $\Delta t$}
\State $\vec{r} \gets \vec{r} + \vec{v} \cdot \Delta t$
\State \Return $\vec{r}$
\EndFunction
\\
\Function{Tumble}{$\vec{v}$, $\Delta t$}
\State $\theta \sim \Gamma(\text{shape}=4, \text{scale}=18.32)$
\LineComment{Sample tumbling angle from distribution given in \cite{sneddon_stochastic_2012}.}

\State $\vec{v} \gets R(\theta) \cdot \vec{v}$
\LineComment{$R(\theta)$ is a 2D rotation matrix through angle $\theta$.}

\State \Return $\vec{v}$
\EndFunction
\\
\Function{OU-Euler-Maruyama}{$m$, $L$, $\Delta t$}
\State $\bar{m} \gets$ \Call{MeanMeth}{$L$, $mb_0$, $\alpha$}
\LineComment{Mean methylation level $\bar{m}(L, mb_0, \alpha)$ as in \cite{sneddon_stochastic_2012,frankel_adaptability_2014}.}
\State $m \gets m + \big[\Delta t/\tau(\bar{m} - m) + \sigma_m\sqrt{2/\tau}dW(\Delta t)\big]$
\State \Return $m$
\EndFunction
\\
\Procedure{SimulateFineEcoliCell}{$t_\text{end}$}
\State $t\gets0$
\While {$t< t_\text{end}$}
    \State $L\gets L(\vec{r}, t)$
    \LineComment{The ligand field $L$ value, at the cell's location $\vec{r}$.}
    \State $\mathbf{s} \gets$ \Call{pCTMC}{$\mathbf{s}$, $m$, $L$, $\Delta t$}
    	\label{st:pctmc}
    	\LineComment{Drawing F/M pCTMC trajectory of length $\Delta t$, with parameters $k_\pm(m, L)$ and initial state the last pCTMC state of the cell.}
    \State $\psi \gets \phi_\text{RUN}(\mathbf{s})$
    \LineComment{Evaluating the $\phi_\text{RUN}$ on (the final state of) the pCTMC trajectory.}\label{st:phi}
    \If{$\psi$}
    	\State $\vec{r} \gets$ \Call{Run}{$\vec{r}$, $\vec{v}$, $\Delta t$}
    \Else
    	\State $\vec{v} \gets$ \Call{Tumble}{$\vec{v}$, $\Delta t$}
    \EndIf
    \State $m \gets$ \Call{OU-Euler-Maruyama}{$m$, $L$, $\Delta t$} \Comment{Evolving methylation.}
    \State $t \gets t + \Delta t$
\EndWhile
\EndProcedure
\end{algorithmic}
\end{algorithm}

\section{Methodology for statistical abstraction}\label{sec:method}

{
In a multi-scale system, output from a set of processes in one layer in the system is passed as input to another layer; these processes are often computationally expensive. We present a methodology to abstract away such a set of processes and replace them with a more efficient stochastic map from the input to the output, governed by an underlying probability function. We approximate this probability function using Gaussian processes after observing many input-output pairs from the processes to be abstracted. The output consists of truth evaluations of properties expressed in logical formulae, which capture some behaviour of the system that is to be preserved by the abstraction.

}

\subsection{Statistical abstraction framework}\label{sec:method:highlevel}

Consider a CTMC $S$, which given an initial state $\mathbf{s}_0$, running time $t$, and input $\mathbf{q}$ which completely determines transition rates, generates a trajectory $\mathbf{s}_{[0, t]}$. The trajectory is then checked for satisfaction of a property resulting in output $y = f(\mathbf{s}_{[0, t]}), y\in\{\top, \bot\}$. This layer of the multi-scale system can therefore be described as a set of operations:
\begin{align}
S(\mathbf{s}_0, t, \mathbf{q}) &= \mathbf{s}_{[0, t]}; \label{eq:fine_layer:S} \\
f(\mathbf{s}_{[0, t]}) &= y. \label{eq:fine_layer:f}
\end{align}
Note that we consider a single property here for simplicity so a single binary value, but one could generalise to multiple properties, and hence, multi-valued output. This output then becomes input to a higher layer in the multi-scale system.

Our goal is to construct a system $\tilde{S}$ that is cheaper to simulate, whose output will be consistent with the original system $S$. Since the system is stochastic, in this context \emph{consistent} refers to having the same probability distribution for the output random variable $y$. This abstracted system should generate output $y'$ given the last output of the system $y$ and the same input $k$ as before:
\begin{align}
\label{eq:coarse_layer}
\tilde{S}(y, \mathbf{q}) = y'.
\end{align}
Replacing the initial state $\mathbf{s}_0$ input with the previous output $y$ allows us to substitute the whole layer of fine operations (\ref{eq:fine_layer:S}, \ref{eq:fine_layer:f}) with the cheaper abstracted system $\tilde{S}$ (\ref{eq:coarse_layer}), in the multi-scale system. { We regard this abstracted system to be a stochastic map from the internal state of the system and some given input to an output; we then use Gaussian processes to estimate the underlying probability function $\Psi(y, \mathbf{q})$ which governs the output of this stochastic map over the input domain.}

\paragraph{Abstracting the \emph{E.\ coli} chemotaxis pathway}
Returning to our model of the \emph{E.\ coli} chemotaxis pathway, we associate the original system $S$ with the pCTMC system of F/M conformations (Equations \ref{eq:pCTMC1}), along with the OU methylation process in Equation \ref{eq:meth}. The input starting state $\mathbf{s}_0$ is the last F/M state of the pCTMC, and the last methylation level $m$. The simulation time $T$ is the variable $\Delta t$ from Section \ref{sec:bg:chemotaxis}, also used for the integration step-size of the OU in the Euler-Maruyama scheme. The transition rates $k_\pm$ are calculated using the variables $m$ and $L$, the last methylation level and external ligand concentration at the position of the cell, respectively. The output of this system, $\mathbf{s}_t$, is then a sampled pCTMC trajectory and new methylation level. Finally, the run property (\ref{eq:phi_run}) is evaluated on (the last state of) the drawn pCTMC trajectory and the output determines whether the cell `runs' or `tumbles'.

\begin{figure}[tb]
	\centering
	\includegraphics[width=0.7\textwidth]{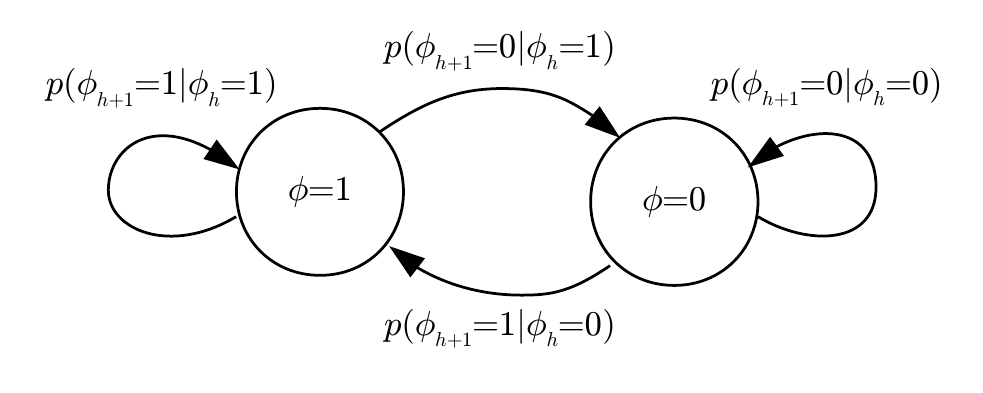}
	\caption{DTMC with two states, $\phi_\text{RUN}\in\{\top,~\bot\}$. The transition probabilities depend on internal methylation level $m$ and external ligand concentration $L$.}
	\label{fig:dtmc}
	\vspace{-1em}
\end{figure}

In observing the truth value of property $\phi_\text{RUN}$ for the state of the pCTMC at regular intervals of $\Delta t$, we cast the original pCTMC model ($S$) into a DTMC (Figure \ref{fig:dtmc}). This DTMC has only two states, $\phi_\text{RUN} \in \{\top, \bot\}$, and transition probabilities depending on the transition rates $k_\pm,~\mu$, of the original pCTMC.

Since this is only a two-state DTMC, the state at the next time-step conditioned on the current one can be modelled as a Bernoulli random variable:
\begin{align}\label{eq:DTMC_bernoulli}
\phi' \mid \phi \sim \text{Bernoulli}(p=p_{\phi'=1\mid\phi} (m, L)),
\end{align}
where $\phi,~\phi'$ are the $\phi_\text{RUN}$ DTMC states at time-steps $h,~h+1$ respectively. Also, the boolean $\{\bot, \top\}$ truth values of the properties have been mapped to the standard corresponding integers $\{0, 1\}$ for mathematical ease.

\begin{algorithm}[!tbp]
\caption{Simulation scheme for the abstracted \emph{E.\ coli} model, based on GP approximation for the RUN/TUMBLE probability. Steps \ref{st:gp}, \ref{st:bern} here replace the expensive Steps \ref{st:pctmc}, \ref{st:phi} in Algorithm \ref{alg:sim-fine}.}
\label{alg:sim-abstract}
\begin{algorithmic}[1]
\Procedure{SimulateAbstractedEcoliCell}{$t_\text{end}$}
\State $t\gets0$
\While {$t< t_\text{end}$}
    \State $L\gets L(\vec{r}, t)$
    \State $p \gets$ \Call{$\text{GP}_\psi$}{$m$, $L$} \label{st:gp}
    \State $\psi \sim \text{Bernoulli}(p)$	\label{st:bern}
    \If{$\psi$}
    	\State $\vec{r} \gets$ \Call{Run}{$\vec{r}$, $\vec{v}$, $\Delta t$}
    \Else
    	\State $\vec{v} \gets$ \Call{Tumble}{$\vec{v}$, $\Delta t$}
    \EndIf
    \State $m \gets$ \Call{OU-Euler-Maruyama}{$m$, $L$, $\Delta t$}
    \State $t \gets t + \Delta t$
\EndWhile
\EndProcedure
\end{algorithmic}
\end{algorithm}

{
We recognise that a \emph{single step} transition of this DTMC ($\phi' \mid \phi, m, L$) is the output $y'\mid y, \mathbf{q}$ produced by the abstracted layer $\tilde{S}(y, \mathbf{q})$. Identifying the corresponding probability function $p_{\phi'=1\mid\phi} (m, L)$ as the underlying governing function $\Psi(y, \mathbf{q})$ completes the setting of \emph{E.\ coli} chemotaxis model abstraction to the methodology framework given above (Section \ref{sec:method:highlevel}). Note that the OU process for methylation is retained in the abstracted model as a parallel running process in the same layer of the multi-scale system. The OU process output $m$, together with the ligand concentration $L$ (output of a different layer in the multi-scale system), constitute the input $\mathbf{q}$. The altered simulation scheme for this abstracted model is outlined in Algorithm \ref{alg:sim-abstract}. Notice how Steps \ref{st:gp}, \ref{st:bern} there replace the more expensive Steps \ref{st:pctmc}, \ref{st:phi} in Algorithm \ref{alg:sim-fine}.
}

\subsection{Approximating the underlying probability function $\Psi$}\label{sec:method:gps}
{
We use Gaussian process (GP) regression in order to infer the underlying probability function $\Psi(y, \mathbf{q})$ governing the stochastic $\tilde{S}$ mapping from internal state $y$ and input $k$ to output $y'$ in Section \ref{sec:method:highlevel}. A GP models a normally distributed stochastic variable over a continuous domain. It can be thought of as a multivariate normal distribution over functions.
This multivariate normal distribution can be conditioned on a finite number of (potentially noisy) observations of the function to be inferred, learning new mean and covariance parameters. These are computable at any point in the domain and correspond to the expected value of the function and associated variance at that point, respectively.}

GPs are universal function approximators. The choice of covariance kernel determines the prior over the function and thus how many observations are required to get a good estimate of the underlying function. However, given enough observations, a GP with any valid kernel will approximate any smooth function arbitrarily well. We refer to \cite{rasmussen_gaussian_2006} for a more comprehensive account of GPs.

{
Since training observations are binary samples of a Bernoulli distribution but GPs regress over a continuous unbounded variable, some adjustments must be made for correct evaluation of the underlying probability function $\Psi$.
These are explained in the Gaussian process classification (GPC) method outlined in \cite{rasmussen_gaussian_2006}, and amount to identifying that class probability function with $\Psi$. We use Minka's Expectation-Propagation (EP) technique to approximate the posterior because it is more accurate than the Laplace approximation. Further, we use fully independent training conditional (FITC) approximation \cite{snelson_sparse_2006} to allow a large number of observations to be considered for learning the underlying function, while maintaining a low cost of predicting at any point of the domain. Note that the Bernoulli distribution likelihood, used here for GPC, is a special case result because of both the binary $y=\phi$ output and the single observation of transitions at a particular $(m,L)$ parametrisation.\footnote{It is highly unlikely to have more than a single transition since $(m, L)$ are continuous values that constantly change for the bacterium.} Lifting these restrictions would result in the more general multinomial distribution. 
}

\paragraph{Constructing $\Psi$ in \emph{E.\ coli} chemotaxis}
{
As we mentioned, a single DTMC transition ($\phi' \mid \phi, m, L$) corresponds to the output $y'\mid y, \mathbf{q}$ produced by the stochastic mapping $\tilde{S}(y, \mathbf{q})$.
Therefore, $\tilde{S}(\phi, (m, L))$ consists of sampling from a Bernoulli distribution $\text{Bernoulli}(p=p_{\phi'=1\mid\phi} (m, L))$ where $p_{\phi'=1\mid\phi} (m, L)$ is the underlying probability function $\Psi(y=\phi,~\mathbf{q}=(m,L))$ in the general formalism.
We approximate $\Psi(y, \mathbf{q}) = p_{\phi'=1\mid\phi} (m, L)$, using GPs trained on observations from \emph{micro-trajectories}, i.e.\ trajectories of the fine F/M pCTMC system which are then mapped onto the property space, $\phi \in \{0, 1\}$, to serve as training data.
}

Therefore, at a given $(m, L)$ the pCTMC with transition rates $k_\pm(m, L)$ is at a state $\mathbf{s}_0$ which maps onto $\phi(\mathbf{s}_0)$. After a time $\Delta t$, the same CTMC is found at a state $\mathbf{s}_{\Delta t}$, which maps onto $\phi(\mathbf{s}_{\Delta t})$. An observation $\phi(\mathbf{s}_{\Delta t}) \mid \phi(\mathbf{s}_0), m, L$ is in this way recorded for every parametrisation $(m, L)$ the bacterium has visited in the micro-trajectories.

Since the output of $\tilde{S}$ is binary ($y=\phi \in \{0, 1\}$) we construct two probability functions $\Psi_\phi(m,L) = p_{\phi'=1|\phi}(m, L)$. Each is approximated with a separate GPC function, where $\Psi_0(m,L)$ is trained on observations of transitions originating from the `TUMBLE' state ($p_{\phi'=1|\phi = 0}(m, L)$) and $\Psi_1(m,L)$ using transitions from the `RUN' state ($p_{\phi'=1|\phi = 1}(m, L)$). Notice that we need not estimate separate functions for $\phi' = \{0, 1\}$, since $p_{\phi' = 1|\phi}(m, L) = 1 - p_{\phi' = 0|\phi}(m, L)$. Having access to these underlying probability functions we are now able to sample the DTMC at any parametrisation $(m, L)$ the bacterium finds itself in, by using the function estimate for $p_\phi(m,L)$ despite not having observations at that $m,L$.

\begin{figure}[htb]
	\centering
		\includegraphics[width=0.43\textwidth]{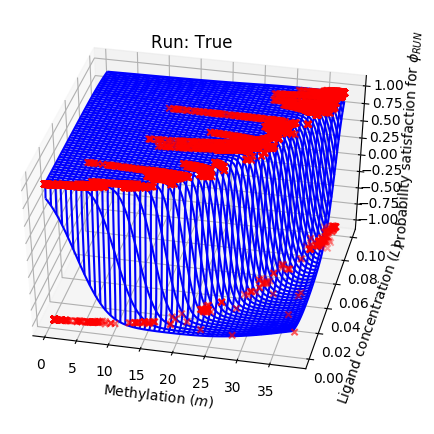}
		\hfill
		\includegraphics[width=0.43\textwidth]{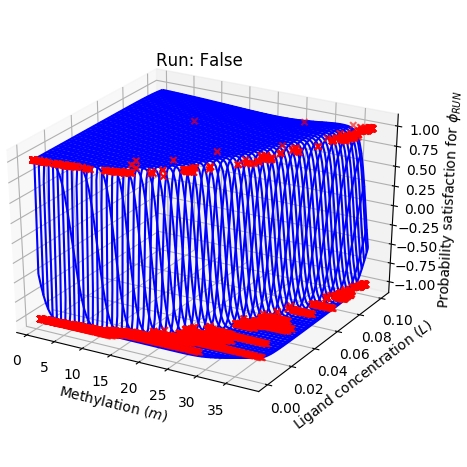}
	\caption{The probability functions $p_{\phi'=1|\phi}(m, L)$ (left: $\phi=\top$, right: $\phi=\bot$) produced by the GP with hyperparameters $\ln (\ell) = (3.5, -2.5)$ and $\ln(\sigma) = 5$, 100 inducing points (FITC approximation), and 10000 observations (red crosses). The steep boundary is accurately captured producing a sharp switch-like transition from the run domain to the tumble domain.}
	\label{fig:gps}
\end{figure}

The function $p_{\phi'=1|\phi}(m, L)$ is particularly challenging for GPs. This is due to a sharp boundary in the $m,~L$ domain, where there is a transition from $p_{\phi'=1|\phi}(m, L)\\\approx0$ to $p_{\phi'=1|\phi}(m, L) \approx 1$. The bacterium has a steady state very close to this boundary, determined by the motor bias $mb_0$, and that is where they are most often found. Therefore, accurate estimation of this boundary is crucial for this problem. Furthermore, the low probability of finding bacteria away from the boundary (in a relatively smooth ligand field) gives a very narrow window of where the function is observed. To get a better overall estimate, we sporadically perturb the position of bacteria in the micro-trajectory phase of collecting observations, such that the bacterium finds itself producing observations away from the boundary for a while, before the system returns close to steady state again. Despite these difficulties, we produce a good reconstruction of the underlying functions $p_{\phi'=1|\phi=0}$ and $p_{\phi'=1|\phi=1}$ over the $m,~L$ domain (see Figure~\ref{fig:gps}).

\section{Results}\label{sec:results}

When assessing performance of our method for statistical abstraction, there are two things of interest: accuracy and computational savings. Accuracy refers to how similar behaviour of the abstracted system is to the behaviour of the original system. In our case of chemotaxis in \emph{E.\ coli}, this is seen by comparing population distributions in a ligand field, resulting from simulations using the original fine system and the abstracted one. We also compare run and tumble duration distributions as another metric of how closely we approximate the output and behaviour of the original model.

Learning the transition probability functions for the dual-state DTMC enabled us to simulate bacteria using our abstracted model on a host of different ligand field profiles. Beyond comparing bacteria population distributions under the original Gaussian ligand field used for learning (see $L_1$ below), we did the same for a linear and dynamic field ($L_2,~L_3$ below), using the same learned functions $p_{\phi'=1|\phi}(m, L),~ \phi\in\{0, 1\}$.

The ligand fields tested were:
\begin{align}
L_1(\vec{r}) &= 0.1\cdot \exp\left[{-0.5(\vec{r}^\top \mathbf{\Sigma}^{-1} \vec{r})}\right],
&&\mathbf{\Sigma} = 3 \cdot \mathbf{I}_2; \\
L_2(\vec{r}) &= \max\bigg(10^{-5},~0.1 - 0.05\sqrt{(\mathbf{A} \vec{r})^\top \mathbf{A} \vec{r}}\bigg),
&&\mathbf{A} = \begin{pmatrix}	1/5 & 0 \\ 0 & 1/2 \end{pmatrix}; \\
L_3(\vec{r}, t) &= 0.1\cdot \exp\left[{-0.5(\vec{r}^\top \mathbf{\Sigma}(t)^{-1} \vec{r})}\right],
&&\mathbf{\Sigma}(t) = 3(t/50 + 1) \cdot \mathbf{I}_2.
\end{align}
In the fields above, the maximum value is 0.1 (units are mM) and this peak concentration is at $\vec{r} = (0, 0)$. The field $L_2$ is a static, non-isotropic, linear field, whereas $L_3$ is a dynamic field: a Gaussian spreading out over time, similar to what one might expect to be produced by a diffusing drop of nutrients.  As expected, as long as the stimulus concentrations and their spatial gradients are within the region observed in training, the population distributions show consistency with those produced when simulating using the original full model. 

\paragraph{Computational cost savings}
Computational savings are given empirically here by comparing running times of simulations for both systems. A hundred (100) cells are simulated in each of the ligand fields, for a time $t_\text{end}=500$s and a time-step of $\Delta t = 0.05$. Therefore, one million (1000000) iterations of the main while loop in Alorithms \ref{alg:sim-fine}, \ref{alg:sim-abstract} are compared in the reported speed-up factor (Table \ref{tab:KS}). We observe a speed-up factor of $\sim8$, reducing running times from $\sim 460\text{m}$ to $\sim 60\text{m}$. Table \ref{tab:KS} reports speed-up factors for each ligand field experiment.

The reported factor values do not include the costs paid for training the GP and producing the training data. It takes $\sim4$m to train GPs for both $\Psi_\phi$ functions, and $\sim10$m for producing 20000 observations of pCTMC transitions from the original fine system (10000 training points for each $\Psi_\phi$ function). The relatively low times compared to simulation times, combined with the fact that one only pays this once, upfront, make these costs negligible.

\paragraph{Accuracy evaluation}
To evaluate how closely results from the abstracted model are compared to the original one, we applied the Kolmogorov-Smirnov (KS) two-sample test \cite{chakravarty_handbook_1967} to the population distributions of the two models at several time-points in the simulation, as well as to the distributions of running and tumbling duration. We have 100 samples from each population distribution since we simulated 100 cells. However, in the case of `Run' and `Tumble' duration distributions we have $\sim60000$ observations from each, because we aggregate observations from the entire trajectory; we choose a random $1000$ sample of these to perform the KS test.\footnote{\label{foot:KS}
	 { We sub-sample because the KS test p-value depends heavily on sample size. Even if two distributions generating samples might be very close, in the limit of an infinite sample size one approaches the true distributions. In such a case, the KS test will reject that the two samples were produced by \emph{the same} distribution, returning lower p-values as sample size increases (for the same KS distance). We do not expect to produce the same distributions here since we are making approximations, so comparing p-values for very large sample sizes is not of interest.}
} In light of these difficulties, a different test which quantifies the distance between the two distributions (e.g. Jensen-Shannon divergence) might be more useful here, but that requires analytic forms of the distributions.


Inspecting Table \ref{tab:KS} we find no KS distance higher than 0.2 indicating very similar distributions, as supported by the associated high p-values. The latter do not allow rejecting the null hypothesis with the current sample, which is that the samples originate from the same distribution. An exception is the `Tumble' duration distributions in the $L_1$ ligand field, where the somewhat higher KS distance of the large sample sizes gives an exaggerated p-value (see footnote \ref{foot:KS}).

We note how even in the case of the dynamic $L_3$ field, the resulting population behaviour of the abstracted model is preserved without any additional training necessary. The fact that the original training occurred in a static field does not affect the ability of the abstract model to cope with a dynamic one.

\begin{table}[htb]
	\caption{KS two-sample test statistics, where the first (top) value reports KS distance and the second (in brackets, bottom) the associated p-value. One sample came from 100 trajectories of fine \emph{E.\ coli} system simulations, and the other from 100 abstracted system simulations. The first four columns show KS test results of original and abstracted bacterial population distances from peak concentration at various times $t$ (shown in Figure \ref{fig:gaussian:dist}). `Run' and `Tumble' columns compare the distributions of run and tumble durations respectively for 1000 samples from each system. The last column reports the observed speed-up factor based on running times and normalising for core utilisation.}
	\label{tab:KS}
	\centering
    
    \setlength\tabcolsep{0.2em}
	\begin{tabular}{l|cccc|c|c|c}
	\hline

	\hline
	\textbf{Field} &
	$t=125\text{s}$ & $t=250\text{s}$ & $t=375\text{s}$ & $t=500\text{s}$ &
	\textbf{Run} & \textbf{Tumble} & \textbf{Speed-up}\\
	\hline

	\hline
		Gaussian: &
		0.110 & 0.160 & 0.170 & 0.160 &
		0.039 & 0.101 &\\
		$L_1(\vec{r})$ &
		(0.556) & (0.140) & (0.099) & (0.140) &
		(0.425) & ($7\cdot10^{-5}$) &
		7.8 \\
	\hline
		Linear: &
		0.010 & 0.150 & 0.170 & 0.130 &
		0.022 & 0.014 &\\
		$L_2(\vec{r})$ &
		(0.677) & (0.193) & (0.100) & (0.344) &
		(0.967) & (0.100) &
		9.4 \\
	\hline
		Dynamic &
		0.140 & 0.070 & 0.140 & 0.080 &
		0.047 & 0.039 &\\
		Gaussian: $L_3(\vec{r}, t)$ &
		(0.261) & (0.961) & (0.261) & (0.894) &
		(0.214) & (0.425) &
		8.9 \\
	\hline

	\hline
	\end{tabular}
\end{table}

\begin{figure}[htb]
	\centering
	\includegraphics[width=0.45\textwidth]{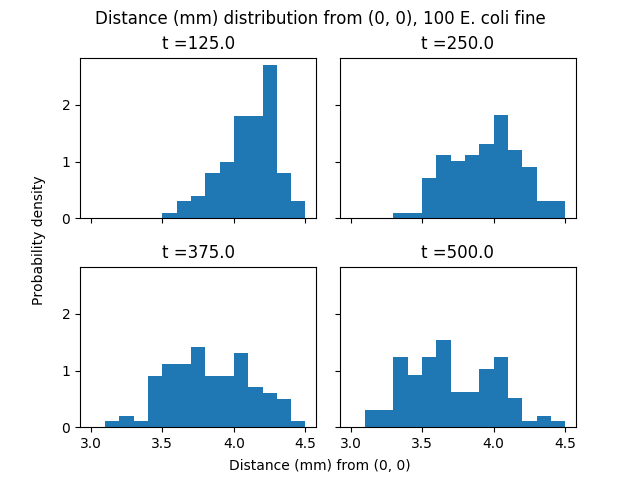}
	~
	\includegraphics[width=0.45\textwidth]{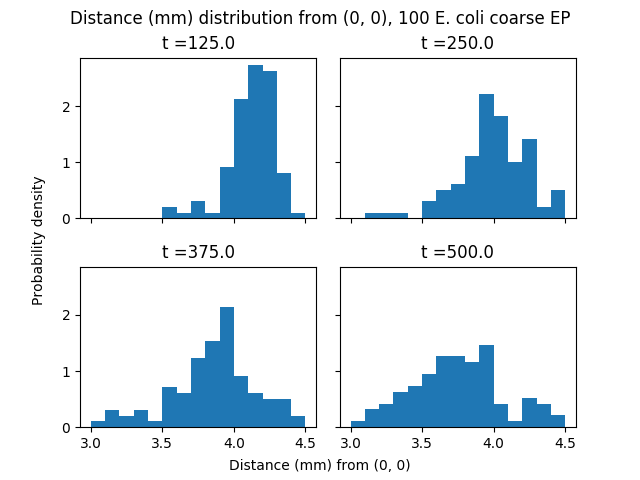}
	\caption{Empirical distributions for the distance of bacteria populations (100 \emph{E.\ coli}) at different times $t$ of the simulation. Left: original full system simulations. Right: abstracted system simulations. Gaussian $L_1$ ligand field.}
	\label{fig:gaussian:dist}
\end{figure}

\begin{figure}[!htb]
	\centering
	\includegraphics[width=0.45\textwidth]{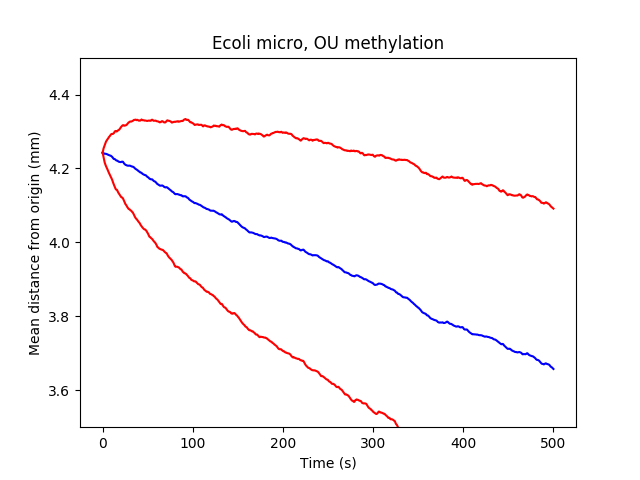}
	~
	\includegraphics[width=0.45\textwidth]{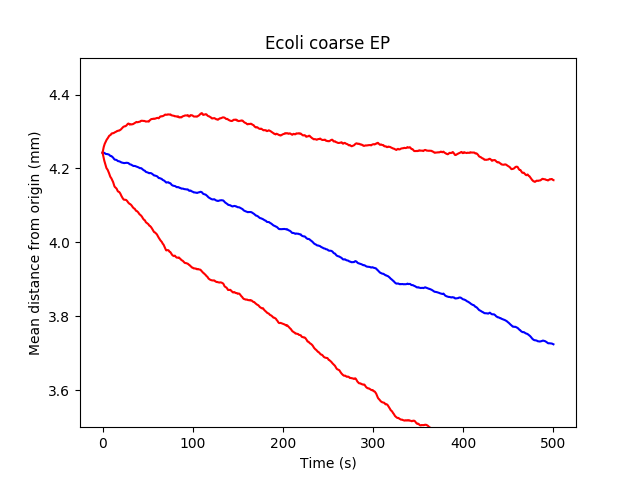}
\\
	\includegraphics[width=0.45\textwidth]{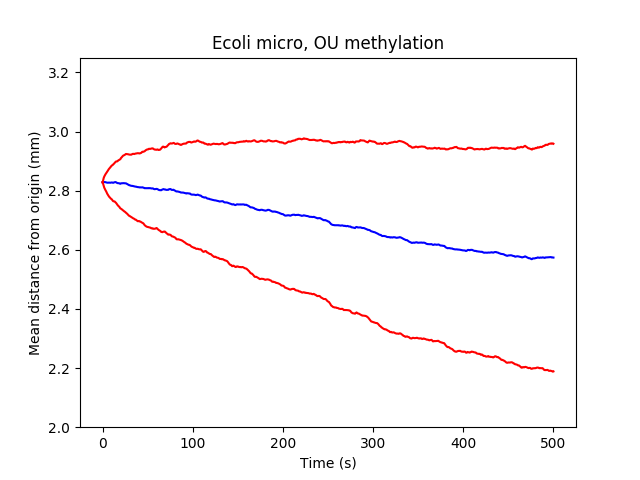}
	~
	\includegraphics[width=0.45\textwidth]{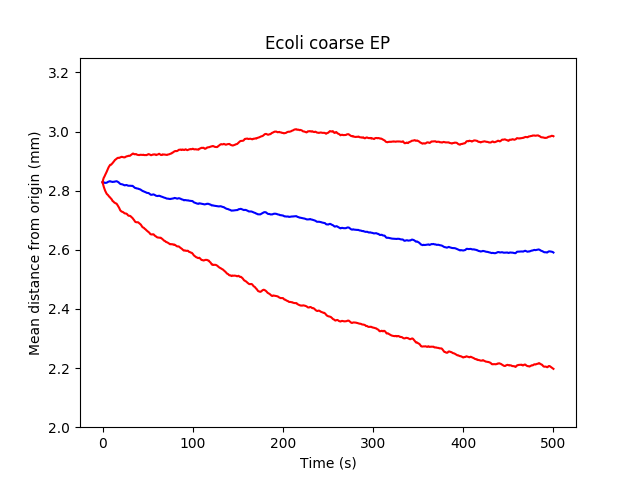}
\\
	\includegraphics[width=0.45\textwidth]{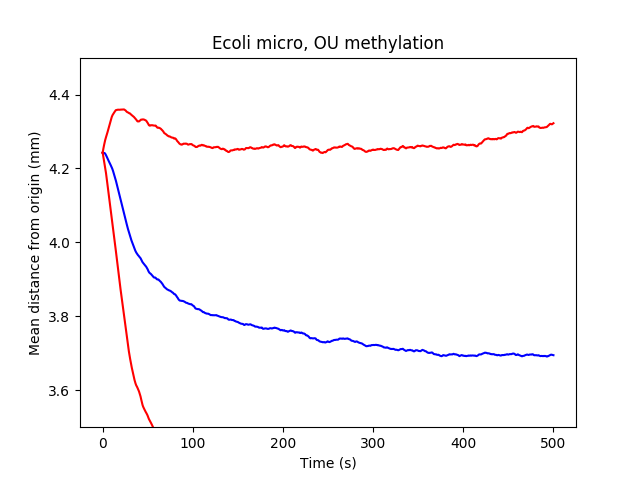}
	~
	\includegraphics[width=0.45\textwidth]{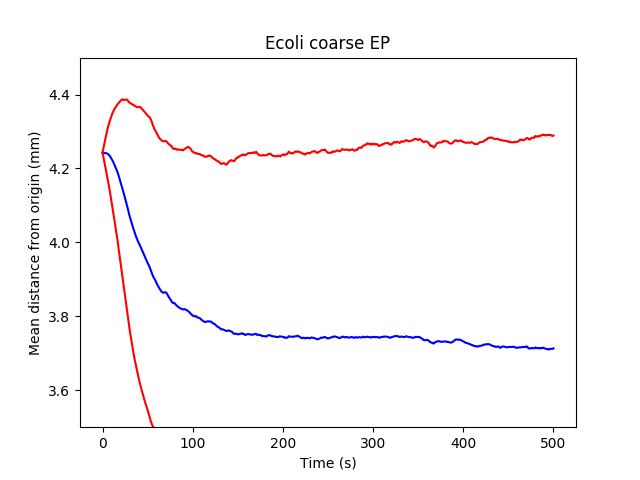}
	\caption{Average (blue) and standard deviation (red) of distance from peak ligand concentration for a population of 100 \emph{E.\ coli} over a time of 500s. Left: original full system simulations. Right: abstracted system simulations. Rows (top to bottom): $L_1, L_2, L_3$ ligand fields respectively.}
	\label{fig:dynamicgaussian:traj}
\end{figure}



\section{Discussion}\label{sec:conclusion}

In many domains, ranging from cyber-physical systems to  biological  and medical processes, consideration of spatio-temporal aspects of behaviour is essential.  However, this comes at great computational expense.  We have presented a methodology that allows layers of a computationally intensive multi-scale model to be replaced by more efficient abstract representations. This is a stochastic map, constructed based on some exploratory simulations of the full model and GP regression.  Our results show that we are able to achieve significant speed-up without sacrificing accuracy. This establishes a framework for such statistical abstraction on which we plan to elaborate in future work.

It should be noted that the specifics of the abstraction are not automatically determined by this framework, but are left to the researcher. Having to manually specify the abstraction introduces an element of flexibility, since different abstractions may be tested and so one can see which are suitable and produce accurate approximations, indicating that pertinent elements of the original model are preserved in the coarsening. Additionally, there may be various valid ways to coarsen a model, depending on what the focus of the inquiry is. On the other hand, it shifts some of the burden of abstracting the model to the researcher, who has to find a suitable set of properties which capture the output behaviour of the layer to be abstracted.

Future work avenues include, for example, allowing more properties to be expressed and using them to guide the abstraction will capture more complex behaviours. Additionally, we could infer abstracted model parameters or underlying functions from real data, instead of synthetic ones. Finally, one would like to be able to deal with correlated agents which result in emergent behaviour at the whole population level. This may be readily achieved in this framework if the interaction between agents happens by altering their modelled external environment (e.g. by manipulating the nutrient field, or by exuding different chemical trails which can be modelled by an additional external field). However, the path is not so clear if the agents are coupled in some other way, where the internal state of one directly affects that of another.

\clearpage

\bibliographystyle{abbrv}
\bibliography{chemotaxis,Markov_chains}

\end{document}